\documentclass[aps,pra,twocolumn,showpacs,floatfix,superscriptaddress]{revtex4-1}
\usepackage{amsmath,amsxtra,amssymb,latexsym,amscd,amsthm}
\usepackage{indentfirst}
\usepackage[mathscr]{eucal}
\usepackage{graphicx}
\usepackage{natbib}
\usepackage{color}

\newcommand{\Tr}{\mathop{\mathrm{Tr}}\nolimits}

\begin{document}
\title{Dissipation in a superconducting artificial atom due to a single non-equilibrium quasiparticle}

\author{D. V. Nguyen}
\affiliation{
Laboratoire de Physique et Mod\'elisation des Milieux Condens\'es,
Universit\'e Grenoble Alpes and CNRS,
25 rue des Martyrs, 38042 Grenoble, France}
\affiliation{Institut N\'eel, Universit\'e Grenoble Alpes and CNRS,
25 rue des Martyrs, 38042 Grenoble, France}

\author{G. Catelani}
\affiliation{
JARA Institute for Quantum Information (PGI-11),
Forschungszentrum J\"ulich, 52425 J\"ulich, Germany}

\author{D. M. Basko}
\affiliation{
Laboratoire de Physique et Mod\'elisation des Milieux Condens\'es,
Universit\'e Grenoble Alpes and CNRS,
25 rue des Martyrs, 38042 Grenoble, France}

\begin{abstract}
We study a superconducting artificial atom which is represented by a single Josephson junction or a Josephson junction chain, capacitively coupled to a coherently driven transmission line, and which contains exactly one residual quasiparticle (or up to one quasiparticle per island in a chain). We study the dissipation in the atom induced by the quasiparticle tunneling, taking into account the quasiparticle heating by the drive. We calculate the transmission coefficient in the transmission line for drive frequencies near resonance and show that, when the artificial atom spectrum is nearly harmonic, the intrinsic quality factor of the resonance increases with the drive power. This counterintuitive behavior is due to the energy dependence of the quasiparticle density of states.
\end{abstract}
\pacs{85.25.Cp, 74.50.+r, 74.81.Fa}

\maketitle

\section{Introduction}

Quantum engineering in superconducting nanocircuits is a rapidly developing field, thanks to progress in sample fabrication techniques which has been occurring in the past decade~\cite{Jung2014}. Due to superconductivity, electromagnetic signals propagate in such circuits with extremely low losses, and the circuit properties can be tuned by applying an external magnetic field. Using superconducting circuit technology, a single microwave photon can be strongly coupled to an artificial atom represented by a superconducting qubit~\cite{Walraff2004}. An artificial atom (AA) can be probed spectroscopically by coupling it to an open superconducting transmission line (TL) and by measuring resonances in reflection or transmission of TL photons at frequencies corresponding to the transitions between the AA energy levels~\cite{Astafiev2010,Hoi2013}.

The AA transitions are broadened by a variety of mechanisms. By analyzing the resonance shape, one can separate the extrinsic broadening, which arises because of the coupling between the AA and the TL and is essentially due to spontaneous emission of photons into the TL, and intrinsic broadening, which is due to dissipation in the AA itself~\cite{Geerlings2012, Masluk2012}. Here we focus on a specific intrinsic dissipation mechanism, which is due to non-equilibrium quasiparticles. At low temperatures, the quasiparticle density is expected to be very low, determined by thermal activation across the superconducting gap. However, many experiments indicate that residual quasiparticles often remain trapped in the
sample~\cite{deVisser2011,Lenander2011,Rajauria2012,Wenner2013,%
Riste2013,Wang2014}, and their recombination can be extremely slow~\cite{Martinis2009,Bespalov2016}.

Many experiments involving residual quasiparticles are successfully described by the theory developed in Refs.~\cite{Catelani2011a,Catelani2011b}. This theory is based on the assumption of a fixed average quasiparticle distribution which perturbs the superconducting degrees of freedom; the resulting net effect is equivalent to that of a frequency-dependent resistance included in the circuit. Technically, this corresponds to a description in terms of the AA reduced density matrix, while the quasiparticles are treated as a bath whose effect can be accounted for by standard dissipative terms in the master equation. The fixed distribution assumption is valid in the weak signal regime, when the back-action of the superconducting condensate excitations on the quasiparticles can be neglected.
This assumption must be reconsidered in situations when the probing signal is strong enough to modify the quasiparticle distribution and the latter can affect the quantities which are measured.

\begin{figure}
\includegraphics[width=8cm]{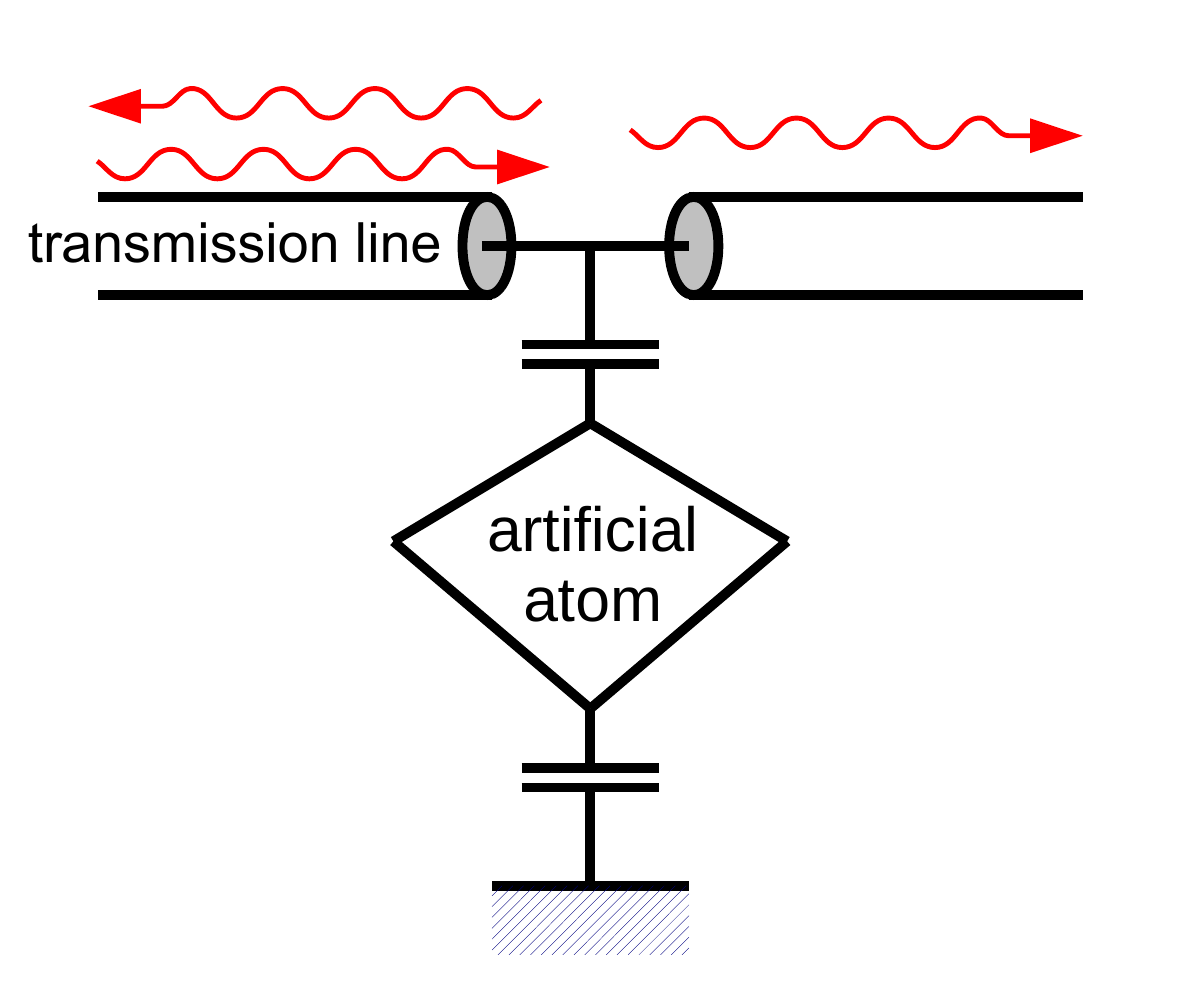}
\caption{\label{fig:transmission} A schematic representation of an artificial atom capacitively coupled to a superconducting transmission line. A coherent signal is sent into the transmission line, whose reflection and transmission are measured.
}
\end{figure}

Here, we study a simple model problem of an AA which is capacitively coupled to a coherently driven TL, as schematically shown in Fig.~\ref{fig:transmission}, and which contains exactly one quasiparticle. Indeed, if the AA initially contains one quasiparticle, it cannot escape into the external circuit because of the capacitors, and has no partner to recombine with. At the same time, we assume the drive to be not too strong, so the system remains at low energy and new quasiparticles are not produced. The AA is represented by a Josephson junction (or a chain of junctions) whose Josephson energy strongly exceeds the Coulomb charging energy.
Technically, we derive the master equation for the AA coupled to a TL analogously to Refs.~\citep{Blais2004,Peropadre2013}, but the quasiparticle degrees of freedom are included in the reduced density matrix following the approach of Refs.~\cite{Breuer,Esposito2003} and its application to a Cooper-pair box in Ref.~\cite{Lutchyn2006}.
Here we focus on the simplest case, assuming the energy exchange with the AA excitations to be the only mechanism of the quasiparticle energy relaxation and fully neglecting acoustic phonon emission by the quasiparticle. When both mechanisms are included, the competition between them results in a variety of different regimes, which will be studied in a forthcoming publication~\cite{Catelani2018}.

Under these assumptions, we calculate here the transmission coefficient in the TL and the intrinsic quality factor of the AA transition, which depend on the coherent drive strength. Indeed, the stronger the drive, the higher is the typical quasiparticle energy, the lower is the quasiparticle density of states, the lower is the probability of quasiparticle tunneling. Thus, the intrinsic quality factor increases with the drive strength (as long as new quasiparticles are not produced). We also extend our calculation to the case when the AA is represented by a Josephson junction chain containing a few quasiparticles (less than one per junction)  whose total number is fixed, and calculate the corresponding intrinsic quality factor of the electromagnetic modes of the chain, obtaining the same power dependence. Such power dependence has been observed in high-quality superconducting resonators~\cite{Macha2010,Megrant2012,Goetz2016} and was attributed to a saturation of two-level systems. The mechanism discussed here may provide an alternative explanation for these observations.

The paper is organized as follows: in the next section, we give a qualitative discussion of the main physical ingredients of our study and of the results. In Sec.~\ref{sec:model} we introduce the theoretical model for the system ``TL + AA + quasiparticle'', include the coherent drive, and give the formal expression for the transmission coefficient in terms of an operator average, when the AA is represented by a single Josephson junction. In Sec.~\ref{sec:master}, we derive the master equation for the junction and the quasiparticle, which is solved in Sec.~\ref{sec:solution}. This study is extended to the case of an AA represented by a Josephson junction chain in Sec.~\ref{sec:chain}. In Sec.~\ref{sec:phonon}, we give a simple estimate of the phonon emission rate to check when it can be neglected. Finally, the conclusions are given in Sec.~\ref{sec:conclusions}.

\section{Qualitative picture}

We study the setup schematically shown in Fig.~\ref{fig:transmission}. The AA is represented by a Josephson junction whose Josephson energy $E_J$ strongly exceeds the Coulomb charging energy, $E_C\equiv{e}^2/(2C_J)$, where $C_J$ is the junction capacitance. The energy of the transition between the AA energy levels is $\hbar\omega_p=\sqrt{8E_JE_C}$, where $\omega_p$ is the junction plasma frequency. If the junction happens to host a quasiparticle (for whatever reason), the quasiparticle cannot be evacuated into the external circuit because of the capacitors, and cannot recombine since the electron number parity is conserved. The AA energy levels are broadened due to (i)~spontaneous emission of TL photons with rate~$\Gamma_\mathrm{tl}$, and (ii)~energy exchange between AA and the quasiparticle with two rates $\Gamma_\mathrm{qp}^+,\Gamma_\mathrm{qp}^-$ for the AA going to the upper/lower energy level, respectively. $\Gamma_\mathrm{tl}$ and $\Gamma_\mathrm{qp}^\pm$ determine the external and internal quality factors of the AA resonance, respectively, and we assume $\Gamma_\mathrm{tl},\Gamma_\mathrm{qp}^\pm\ll\omega_p$.

The microscopic mechanism of energy exchange between the AA, built of the superconducting condensate degrees of freedom in the Josephson junction, and the quasiparticle, is the quasiparticle tunneling between the two sides of the junction.
The rates of quasiparticle tunneling accompanied by excitation or deexcitation of the AA can be estimated from the Fermi Golden Rule as
\begin{eqnarray}
&&\Gamma_\mathrm{qp}^\pm\sim\frac\delta\hbar\,\frac{E_J}\Delta\,
\frac{\hbar\omega_p}{E_J}
\sqrt{\frac\Delta{\max\{T_\mathrm{eff},\hbar\omega_p\}}}.\label{Gammaqppm=}
\end{eqnarray}
%
Here $\delta$ is the normal-state mean level spacing of each island forming the junction (for simplicity, the two islands are assumed to be identical), while $T_\mathrm{eff}$ is the effective quasiparticle temperature, or, equivalently, the typical energy of the quasiparticle counted from the quasiparticle band bottom at~$\Delta$; it depends on the drive strength, as we will show in Sec.~\ref{sec:solution}. The factor $E_J/\Delta$ is of the order of the dimensionless junction conductance junction in the normal state, according to Ambegaokar-Baratoff relation~\cite{Ambegaokar1963}; it appears because $E_J$ is proportional to the square of the single-electron tunneling matrix element.
The factor $\hbar\omega_p/E_J$ originates from the first off-diagonal matrix element of the tunneling Hamiltonian in the AA subspace. The last factor is the quasiparticle density of states, assuming $T_\mathrm{eff},\hbar\omega_p\ll\Delta$ (otherwise, more quasiparticles can be produced, which is not taken into account in the present theory).

Strictly speaking, the use of Fermi Golden Rule requires the energy spectrum of the final states to be continuous, while here we are dealing with discrete spectrum. Indeed, the level spacing~$\delta$ is finite because the islands have a finite volume, and one cannot send $\delta\to{0}$ because then the tunneling rate vanishes (this vanishing is due to the simple fact that $\delta\to{0}$ implies the volume going to infinity, so the quasiparticle just never arrives at the junction). The AA energy levels are also discrete. The use of the Golden Rule is consistent is the energy levels are sufficiently dense, so that the rate exceeds the energy spacing of the final states. This spacing is of the order of $\delta\sqrt{\max\{T_\mathrm{eff},\hbar\omega_p\}/\Delta}$, and it is easy to see from Eq.~(\ref{Gammaqppm=}) that the rates $\Gamma_\mathrm{qp}^\pm$ are always smaller. Thus, to allow the quasiparticle tunneling, the AA levels have to be sufficiently broadened by the photon emission, $\hbar\Gamma_\mathrm{tl}\gtrsim\delta\sqrt{\max\{T_\mathrm{eff},\hbar\omega_p\}/\Delta}$ \cite{Phonons,Elastic}.

Thus, we are forced to consider the situation when the internal quality factor is much higher than the external one, because of the condition $\Gamma_\mathrm{qp}^\pm\ll\Gamma_\mathrm{tl}$. Then, the average degree of excitation of the AA is determined by the balance between the coherent drive and the spontaneous photon emission into the TL. In turn, the quasiparticle effective temperature $T_\mathrm{eff}$ is determined by the AA degree of excitation and is found from the solution of the kinetic equation. Then, $T_\mathrm{eff}$ determines $\Gamma_\mathrm{qp}^\pm$ and the internal quality factor.

Since we assumed $E_C\ll{E}_J$, the lower part of the AA energy spectrum corresponds to a weakly anharmonic oscillator, in the sense that the anharmonic correction to the energy levels, $\sim{E}_C$, is smaller than the oscillator transition energy, $\hbar\omega_p$. However, two different situations may arise depending on the relation between $E_C$ and~$\Gamma_\mathrm{tl}$. If $E_C\ll\Gamma_\mathrm{tl}$, tuning the drive frequency in resonance with the first transition automatically puts it in resonance with subsequent transitions, so the AA can be treated as a harmonic oscillator (as long as its degree of excitation is not too high). In the opposite case, $E_C\gg\Gamma_\mathrm{tl}$, the second transition is automatically out of resonance; in this case, AA is effectively a two-level system, also known as the transmon qubit~\cite{Koch2007}. Below we present the theory for both cases; however, for realistic values of the system parameters, the AA represented by a single Josephson junction corresponds to the qubit limit.
It turns out that in the qubit limit, the effect of quasiparticle heating is always masked by the power broadening of the AA transition.

The harmonic limit turns out to be relevant for a slightly more complex realization of the AA, a chain of Josephson junctions. Sufficiently long chains have isolated resonances with high quality factors~\cite{Masluk2012}, while the nonlinear correction to the transition frequency scales as the inverse of the number of junction in the chain~\cite{Weissl2015}. The number of quasiparticles should be proportional to number of junctions, but when there is much less than one quasiparticle per junction, the quasiparticles can be treated independently, so the theory developed for one junction is straightforwardly extended on the case of a long chain.

Our main result is that in both cases the intrinsic quality factor due to the quasiparticle tunneling increases with the drive strength (as long as new quasiparticles are not produced). The reason for such behaviour is very simple and general: the stronger the drive, the higher is the typical quasiparticle energy, the lower is the quasiparticle density of states at such energies, so the lower is the probability of quasiparticle tunneling. 

Our calculations are done assuming that the energy exchange with the AA excitations is the main mechanism of the quasiparticle energy relaxation, and fully neglecting acoustic phonon emission by the quasiparticle. The latter is known to quickly slow down for low quasiparticle energies~\cite{Kaplan1976}; a simple estimate for typical parameters shows that the phonon emission rate is indeed smaller than the quasiparticle tunneling rate; however, the inequality is not very strong. Thus, a study including both mechanisms is needed and will be reported elsewhere~\cite{Catelani2018}.

\section{The model}\label{sec:model}
\subsection{The system Hamiltonian}

\begin{figure*}
\includegraphics[width=12cm]{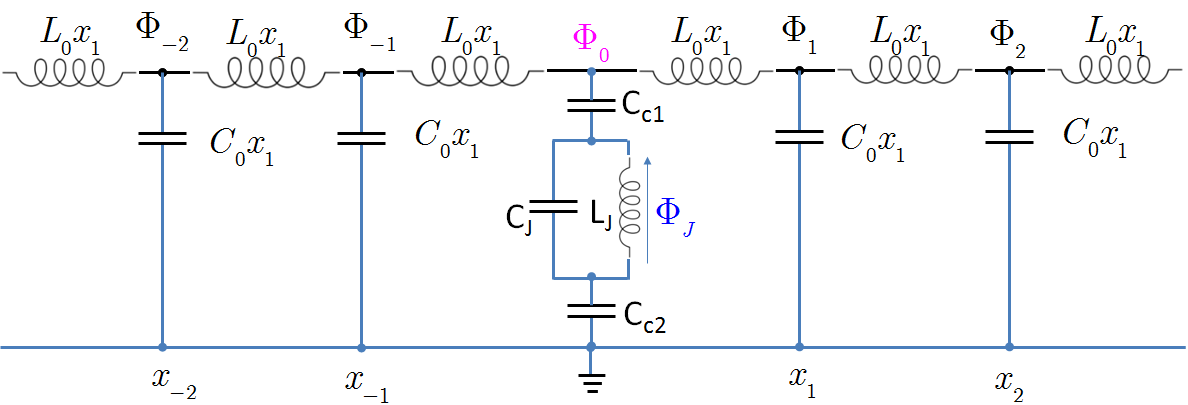}
\caption{\label{fig:FIG1} A schematic circuit representation of a Josephson junction coupled to a transmission line, modeled as an infinite discrete array of inductors and capacitors.
}
\end{figure*}

We consider an artificial atom represented by a single Josephson junction (JJ), made of two superconducting islands, and coupled to a transmission line (TL) by a capacitance $C_{c1}$ and grounded via a capacitance~$C_{c2}$, as shown in Fig. \ref{fig:FIG1}.  The TL is characterized by its inductance $L_0$ and capacitance $C_0$ per unit length, and their ratio determines the TL impedance, $Z_0=\sqrt{L_0/C_0}$. We model the TL by a discrete array of inductors and capacitors with the discretization length~${x_1}$, the limit ${x_1}\to{0}$ to be taken in the end. The JJ is characterized by two energy scales: the Josephson energy $E_J$ and the charging energy $E_C$, related to the Josephson inductance $L_J$ and the junction capacitance $C_J$ as $E_J=(\hbar/2e)^2(1/L_J)$ and $E_C=e^2/(2C_J)$. In the following, we assume $E_J\gg{E}_C$, then the quantum fluctuations of the superconducting phase are small and the JJ can be viewed as a weakly anharmonic oscillator whose linear frequency is the JJ plasma frequency, $\omega_{p}=1/\sqrt{L_JC_J}$.
We assume that the JJ hosts a single quasiparticle which can tunnel between the two islands, but cannot leave the junction because of the capacitors.
This system can be described by the following Hamiltonian
\begin{equation}\label{Htot=}
\hat{H}=\hat{H}_\mathrm{J}+\hat{H}_\mathrm{tl} + \hat{H}_\mathrm{Jtl} + \hat{H}_\mathrm{qp} + \hat{H}_\mathrm{Jqp}.
\end{equation}
The first three terms describe the JJ, the photons in the TL, and their coupling, respectively. They are given by the sum of the electrostatic energy of each capacitor and the energy of each inductor:
\begin{subequations}
\begin{align}
&\hat{H}_\mathrm{J}=\frac{\hat{Q}_J^2}{2C_J}
+ E_J\left(1-\cos\frac{2e\hat\Phi_J}\hbar\right)
\label{Hr=}\\
&\hat{H}_\mathrm{tl}= \frac{\hat{Q}_0^2}{2C_c} + \frac{\hat{Q}_0^2}{2C_J}
+ \sum_{n\neq{0}}\frac{\hat{Q}_n^2}{2C_0{x_1}}
+ \sum_n\frac{(\hat\Phi_n-\hat\Phi_{n+1})^2}{2L_0{x_1}},
\label{HB=}\\
&\hat{H}_\mathrm{Jtl}=\frac{\hat{Q}_J\hat{Q}_0}{C_J}.\label{HrB=}
\end{align}
Here $\hat{Q}_{n\neq{0}}$ is the operator of charge on the upper plate of the $n$th capacitor $C_0{x_1}$ and $\hat\Phi_n$ is the corresponding canonically conjugate flux, $[\hat{Q}_n,\hat\Phi_m]=-i\hbar\delta_{nm}$, whose time derivative is the voltage on the node~$n$. At $n=0$, $\hat\Phi_0$ is related to the voltage of the node~$n=0$, while $\hat\Phi_J$ is related to the voltage drop across the junction. The conjugate charges $\hat{Q}_0$ and $\hat{Q}_J$ are given by the appropriate linear combinations of the charges on the three capacitors $C_{c1}$, $C_{c2}$, and $C_J$. The electrostatic energy of the $n=0$ node is expressed in terms of $C_c=C_{c1}C_{c2}/(C_{c1}+C_{c2})$, the series capacitance of capacitors $C_{c1}$ and $C_{c2}$. The electrostatic energy of the junction is given by $(\hat{Q}_0+\hat{Q}_J)^2/(2C_J)$, and it is split between the three terms $\hat{H}_\mathrm{J}+\hat{H}_\mathrm{tl} + \hat{H}_\mathrm{Jtl}$. The superconducting phase difference on the junction is given by $\hat\phi=(2e/\hbar)\hat\Phi_J$ (we assume $e>0$, so the electron charge is $-e$).

The last two terms in Eq.~(\ref{Htot=}) describe the quasiparticle and its interaction with the superconducting phase difference on the
JJ~\cite{Catelani2011a,Catelani2011b}:
\begin{align}
&\hat{H}_\mathrm{qp} = \sum_{j = \mathrm{u,l}}\sum_\mathbf{p}
\epsilon_\mathbf{p}|j,\mathbf{p}\rangle\langle j,\mathbf{p}|,\\
&\hat{H}_\mathrm{Jqp} = \sum_{\mathbf{p},\mathbf{p}'}
\mathcal{T}_{\mathbf{p}\mathbf{p}'}
\left(u_\mathbf{p}u_{\mathbf{p}'}-v_\mathbf{p}v_{\mathbf{p}'}
e^{2ie\hat\Phi_J/\hbar}\right)
|\mathrm{u},\mathbf{p}\rangle\langle\mathrm{l},\mathbf{p}'| +{}\nonumber\\
&\qquad\qquad{}+\mbox{h.\:c.}
\label{HJqp=}
\end{align}
\end{subequations}
Here $|j,\mathbf{p}\rangle$ is the state of the quasiparticle on the upper/lower island of the junction, $j=\mathrm{u,l}$, with momentum~$\mathbf{p}$. The quasiparticle energies, $\epsilon_\mathbf{p}=\sqrt{\xi_\mathbf{p}^2+\Delta^2}-\Delta$, measured from the gap $\Delta$, are assumed to be the same for both islands. The quasiparticle energy in the normal state, $\xi_\mathbf{p}$, determines the normal state density of states per spin projection, which can also be represented as the inverse of the mean level spacing~$\delta$ on each island:
\begin{equation}
\frac{1}\delta=\sum_\mathbf{p}\delta(\xi_\mathbf{p}-\xi).
\end{equation}
$\delta$~is assumed to be energy-independent. Being inversely proportional to the island volume, $\delta$~is small but finite. The quasiparticle density of states is given by
\begin{equation}\label{nu=}
\nu(\epsilon)\equiv\sum\limits_{\bf{p}} \delta(\epsilon_{\bf{p}} - \epsilon) =
\frac{2}\delta\,\frac{\theta(\epsilon)(\epsilon + \Delta)}%
{\sqrt{(\epsilon+\Delta)^2 - \Delta^2}},
\end{equation}
where $\theta(\epsilon)$ is the Heaviside step function.
The quasiparticle Bogolyubov amplitudes are given by
\begin{equation}
v_\mathbf{p}^2=1-u_\mathbf{p}^2
=\frac{1}2\left(1-\frac{\xi_\mathbf{p}}{\epsilon_\mathbf{p}+\Delta}\right).
\end{equation}
The tunnelling matrix elements, $T_{\mathbf{p}\mathbf{p}'}$,
are assumed to be real, symmetric, and energy-independent,
in which case they are related to the Josephson energy by
the Ambegaokar-Baratoff relation~\cite{Ambegaokar1963}:
\begin{equation}
\sum_{\mathbf{p},\mathbf{p}'}\mathcal{T}_{\mathbf{p}\mathbf{p}'}^2
\delta(\xi_{\mathbf{p}'}-\epsilon')\,\delta(\xi_\mathbf{p}-\epsilon)
=\frac{E_J}{\pi^2\Delta}.
\end{equation}
Below we assume that the quasiparticle energy always remains small, $\epsilon_\mathbf{p}\ll\Delta$, so we approximate $\epsilon_\mathbf{p}\approx\xi_\mathbf{p}^2/(2\Delta)$,  $\nu(\epsilon)\approx (1/\delta)\sqrt{2\Delta /\epsilon}$, and we expand $u_\mathbf{p},{v}_\mathbf{p}\approx
{1}/\sqrt{2}\pm(1/2)\sqrt{\epsilon_\mathbf{p}/\Delta}$.
Also, in the regime of small phase oscillations, we expand $e^{2ie\hat\Phi_J/\hbar}\approx{1}+{2ie\hat\Phi_J/\hbar}$. Then, the matrix element of the tunneling Hamiltonian~(\ref{HJqp=}) becomes
\[
u_\mathbf{p}u_{\mathbf{p}'}-v_\mathbf{p}v_{\mathbf{p}'}
e^{2ie\hat\Phi_J/\hbar}\approx
\frac{\sqrt{\epsilon_\mathbf{p}}+\sqrt{\epsilon_{\mathbf{p}'}}}%
{\sqrt{2\Delta}}
-\frac{ie\hat\Phi_J}\hbar.
\]
The first term of this expression corresponds to elastic quasiparticle tunneling without changing the JJ state; when inserted into the Fermi Golden Rule, it produces the estimate for the tunneling rate given in Ref.~\cite{Elastic}. The second term describes quasiparticle tunneling which induces a transition between the JJ energy levels up or down by one level, and is the crucial ingredient for the master equation, derived below.

\subsection{Coherent drive and transmission coefficient}
\label{ssec:drive}

In the following, we assume that the JJ is probed by sending a coherent wave in the transmission line and measuring its amplitude transmission coefficient $S_{21}$. Our calculation will focus on the dynamics of the JJ degrees of freedom, $\hat{Q}_J$ and $\hat\Phi_J$, so we would like to express the observable $S_{21}$ in terms of the quantum average $\langle\hat{Q}_J\rangle$. To do this, let us write the Heisenberg equations of motion for the TL operators:
\begin{subequations}\begin{align}
&\frac{d\hat{Q}_n}{dt}=
\frac{\hat\Phi_{n+1}+\hat\Phi_{n-1}-2\hat\Phi_n}{L_0{x_1}},\\
&\frac{d\hat\Phi_n}{dt}=\frac{\hat{Q}_n}{C_0{x_1}}\quad(n\neq{0}),\\
&\frac{d\hat\Phi_0}{dt}=\frac{\hat{Q}_0}{C_c}
+\frac{\hat{Q}_0}{C_J}+\frac{\hat{Q}_J}{C_J}.
\end{align}\end{subequations}
These equations are linear, so their solution can be formally written as
\begin{equation}\label{Qsource=}
\hat{Q}_n(t)=\hat{Q}_n^\mathrm{free}(t)
+\int\limits_{-\infty}^tG_n(t-t')\,\hat{Q}_J(t')\,dt',
\end{equation}
where $\hat{Q}_n^\mathrm{free}(t)$ is a solution for the free TL (i.~e., taking into account the Hamiltonian $\hat{H}_\mathrm{tl}$ only), while the last term represents the effect of $\hat{H}_\mathrm{Jtl}$ with $\hat{Q}_J(t)$ treated as a source. $G_n(t-t')$ is the retarded Green's function, given by
\begin{align}
& G_n(t-t')=\int\frac{d\omega}{2\pi}\,
e^{-i\omega(t-t')}G_n(\omega),\\
&G_n(\omega)=
\frac{[1-i\delta_{n0}\cot(k_\omega{x_1}/2)]
C_0{x_1}C_ce^{ik_\omega|n|{x_1}}}%
{C_cC_J-C_0{x_1}(C_J+C_c)[1-i\cot(k_\omega{x_1}/2)]},
\end{align}
where $k_\omega$ is the wave vector, related to $\omega$ by the TL dispersion relation,
\begin{equation}\label{TLdispersion=}
\frac{\omega^2x_1^2}{v^2}=4\sin^2\frac{k_\omega{x_1}}{2},\quad
v^2\equiv\frac{1}{L_0C_0}.
\end{equation}
In the continuum limit, ${x_1}\to{0}$, $n{x_1}=x$,
$G_n(\omega)={x_1}\mathcal{G}(x,\omega)$,
the expressions simplify as $k_\omega=\omega/v$ and
\begin{equation}
\mathcal{G}(x,\omega)=\frac{-i\omega\tau_c(C_0/C_J)}{1+C_c/C_J-i\omega\tau_c}
\left[1-\frac{2iv\delta(x)}\omega\right]e^{i\omega|x|/v},
\end{equation}
where $\tau_c\equiv{C}_cZ_0/2$ is the classical $RC$-time of the $C_c$~capacitor coupled to the TL.

The free part $\hat{Q}_n^\mathrm{free}(t)$ is assumed to be the sum of the vacuum part with zero quantum average and the classical part. The latter contains the incident coherent wave with frequency~$\omega_d$, momentum~$k_d$ determined by the dispersion relation~(\ref{TLdispersion=}), and voltage amplitude $V_d$, as well as the scattered wave:
\begin{align}
&\frac{\langle\hat{Q}_n^\mathrm{free}(t)\rangle}{C_0{x_1}}
=V_de^{-i\omega_dt}\left(e^{ik_dn{x_1}}+re^{ik_d|n|{x_1}}\right)
(1+\zeta\delta_{n0})+{}\nonumber\\
&\qquad\qquad\quad{}+\mathrm{c.\:c.},\label{avQfree=}\\
&r\equiv\frac{i\zeta\tan(k_d{x_1}/2)}{1-i\zeta\tan(k_d{x_1}/2)},\quad
\zeta\equiv\frac{C_cC_J}{(C_c+C_J)C_0{x_1}}-1.
\end{align}
The scattered wave appears in $\hat{Q}_n^\mathrm{free}(t)$ because
$\hat{H}_\mathrm{tl}$ in Eq.~(\ref{HB=}) is not translationally invariant.
Indeed, the $n=0$ site differs from all other sites by the coefficient
at~$\hat{Q}_0^2$. Thus, left- and right-traveling waves are not normal
modes even for the ``free'' TL.
Taking the quantum average of Eq.~(\ref{Qsource=}) and the ratio of the
transmitted wave amplitude to the incident one [note that the last
term in Eq.~(\ref{Qsource=}) does not contain the incident wave],
we can relate the transmission coefficient $S_{21}$ to the average
$\langle\hat{Q}_J(t)\rangle\equiv Q_+e^{-i\omega_dt}+Q_+^*e^{i\omega_dt}$
as
\begin{equation}
\label{S21}
S_{21}(\omega_d)=\frac{1+C_c/C_J-i\omega_d\tau_c{Q}_+/(C_JV_d)}%
{1+C_c/C_J-i\omega_d\tau_c},
\end{equation}
where the continuum limit ${x_1}\to{0}$ has been taken.

In the next section, we will study the master equation for the JJ and the quasiparticle, treating the TL as a bath. It is much simpler to write down the master equation when the bath is in the vacuum state, rather than in a coherent state. Thus, we will replace the above system with the driven TL by an equivalent one, where the TL is in the vacuum state, but the oscillator is driven directly.
To see this equivalence, we write down the Heisenberg equations of motion for $\hat{Q}_J$
and $\hat\Phi_J$:
\begin{subequations}\begin{align}
&\frac{d\hat{Q}_J}{dt}=-\frac{\hbar}{2eL_J}\sin\frac{2e\hat\Phi_J}\hbar
+\frac{i}\hbar[\hat{H}_\mathrm{Jqp},\hat{Q}_J],\\
&\frac{d\hat\Phi_J}{dt}=\frac{\hat{Q}_J}{C_J}+\frac{\hat{Q}_0}{C_J}.
\end{align}\end{subequations}
The JJ is driven by the incident coherent wave via the last term,
$\hat{Q}_0/C_J$. Let us now recall that $\hat{Q}_0$ can be represented in the form~(\ref{Qsource=}) where the first term $\hat{Q}_n^\mathrm{free}$ contains the vacuum part and the coherent part including the incident wave, while the second term in $\hat{Q}_0$ does not contain the incident field. Thus, the Heisenberg equation for $\hat{Q}_J$, $\hat\Phi_J$ will have exactly the same form if we assume $\hat{Q}_n^\mathrm{free}$ to have only vacuum contribution, while $\hat{Q}_J$ is driven by an external voltage $V_J(t)=\langle\hat{Q}_0^\mathrm{free}\rangle/C_J$. In other words, the JJ quantum dynamics is the same if no incident field is sent in the TL, but an additional driving term is introduced in the JJ Hamiltonian:
\begin{equation}\label{Hd=}
\hat{H}_\mathrm{d}=\frac{C_c}{C_J}
\left(\frac{V_de^{-i\omega_dt}}{1+C_c/C_J-i\omega_d\tau_c}
+\mathrm{c.\:c.}\right)\hat{Q}_J.
\end{equation}
The perturbative master equation derived in the next section assumes the weak-coupling limit, that is, $C_c\ll{C}_J$ and $\omega_pC_cZ_0\ll{1}$.
Then, the denominator in the brackets can be set to unity.

\section{Master equation}\label{sec:master}

It is convenient to rewrite the bosonic part of the Hamiltonian in terms of the creation and annihilation operators. For the JJ operators we have the standard harmonic oscillator expressions,
\begin{equation}
\label{quanti}
\hat\Phi_J=\sqrt{\frac{\hbar{Z}_J}{2}}
\left(\hat{a}+\hat{a}^\dagger\right),\quad
\hat{Q}_J=\sqrt{\frac\hbar{2Z_J}}\,
\frac{\hat{a}-\hat{a}^\dagger}{i},
\end{equation}
where the JJ impedance $Z_J=\sqrt{L_J/C_J}$; then the harmonic part of the JJ Hamiltonian becomes $\hbar\omega_p(\hat{a}^\dagger\hat{a}+1/2)$, where the plasma frequency $\omega_{p}=1/\sqrt{L_JC_J}$.
For the TL in the continuum limit, $nx_1\to{x}$, $x_1\to{0}$, we introduce the fields  $\hat\Phi_n\to\hat\Phi(x)$ and $\hat{Q}_n\to{x_1}\hat{q}(x)$, which are expressed in terms of normal modes of the Hamiltonian $\hat{H}_\mathrm{tl}$~(\ref{HB=}). As discussed in Sec.~\ref{ssec:drive}, these normal modes are not given by left- and right-travelling waves, because of scattering at $n=0$. Taking advantage of the symmetry $n\to-n$, we separate the normal modes into even (e) and odd (o), so the flux and charge density fields are represented as
\begin{subequations}\begin{align}
&\hat\Phi(x)=
\int\limits_0^\infty d\omega\,\sqrt{\frac{\hbar{Z_0}}{2\pi\omega}}
\left[
\left(\hat{b}_{\mathrm{e},\omega}+\hat{b}_{\mathrm{e},\omega}^\dagger\right)
\cos\left(\frac{\omega|x|}{v}+\theta_\omega\right)\right.+\nonumber\\
&\qquad\qquad\quad{}+\left.
\left(\hat{b}_{\mathrm{o},\omega}+\hat{b}_{\mathrm{o},\omega}^\dagger\right)
\sin\frac{\omega{x}}{v}\right].
\end{align}
\begin{align}
&\hat{q}(x)=\left[C_0+\frac{C_cC_J}{C_c+C_J}\,\delta(x)\right]
\int\limits_0^\infty d\omega\,\sqrt{\frac{\hbar\omega{Z_0}}{2\pi}}
\times{}\nonumber\\
&\qquad\qquad{}\times\left[
\frac{\hat{b}_{\mathrm{e},\omega}-\hat{b}_{\mathrm{e},\omega}^\dagger}{i}
\cos\left(\frac{\omega|x|}{v}+\theta_\omega\right)\right.+\nonumber\\
&\qquad\qquad\quad{}+\left.
\frac{\hat{b}_{\mathrm{o},\omega}-\hat{b}_{\mathrm{o},\omega}^\dagger}{i}
\sin\frac{\omega{x}}{v}\right].
\end{align}\end{subequations}
Here $\theta_\omega=\arctan[\omega\tau_c/(1+C_c/C_J)]$ is the scattering phase shift, and the commutation relations for the bosonic operators are
\begin{equation}
[\hat{b}_{j\omega},\hat{b}^\dagger_{j'\omega'}]=
\delta_{jj'}\delta(\omega-\omega'),\quad
j,j'=\mathrm{e,o}.
\end{equation}
Note the $\delta(x)$ contribution to $\hat{q}(x)$; it corresponds
to a finite value of $\hat{Q}_0=\int_{0^-}^{0^+}\hat{q}(x)\,dx$.
The resulting Hamiltonian takes the form $\hat{H}_0+\hat{H}_1$,
where
\begin{subequations}\begin{align}
&\hat{H}_0=\int\limits_0^\infty{d}\omega\,\hbar\omega
\left(\hat{b}_{\mathrm{e},\omega}\hat{b}_{\mathrm{e},\omega}^\dagger
+\hat{b}_{\mathrm{o},\omega}\hat{b}_{\mathrm{o},\omega}^\dagger\right)
+{}\nonumber\\ &\qquad{}+
\hat{H}_J
-i\hbar\left(\Omega{e}^{-i\omega_{d}t}+\Omega^*{e}^{i\omega_{d}t}\right)
\left(\hat{a}-\hat{a}^\dagger\right)+{}\nonumber\\ &\qquad{}+
\sum_\mathbf{p}\frac{\xi_\mathbf{p}^2}{2\Delta}
\left(|\mathrm{l},\mathbf{p}\rangle\langle\mathrm{l},\mathbf{p}|
+|\mathrm{u},\mathbf{p}\rangle\langle\mathrm{u},\mathbf{p}|\right),\\
&\hat{H}_1=-
\int\limits_0^\infty{d}\omega\,\hbar\kappa(\omega)
\left(\hat{b}_{\mathrm{e},\omega}-\hat{b}_{\mathrm{e},\omega}^\dagger\right)
\left(\hat{a}-\hat{a}^\dagger\right)
+{}\nonumber\\ &\qquad\quad{}+\sum_{\mathbf{p},\mathbf{p}'}
i\tilde{\mathcal{T}}_{\mathbf{p}\mathbf{p}'}
\left(|\mathrm{l},\mathbf{p}\rangle\langle\mathrm{u},\mathbf{p}'|
-|\mathrm{u},\mathbf{p}\rangle\langle\mathrm{l},\mathbf{p}'|\right)
\left(\hat{a}+\hat{a}^\dagger\right),
\end{align}\end{subequations}
where $\hat{H}_0$ describes the TL photons, the JJ excitations (plasma oscillations) driven by an external force [related to the incident wave amplitude via Eq.~(\ref{Hd=})], and the quasiparticle, while $\hat{H}_1$ describes the coupling between the TL and the JJ, as well as the JJ coupling to the quasiparticle.
The coupling constants for the JJ-TL coupling, the external drive strength, and the JJ-quasiparticle coupling amplitude are given by
\begin{subequations}\begin{align}
&\kappa(\omega)=\sqrt{\frac{\omega{Z}_0}{4\pi{Z}_J}}\,
\frac{C_c}{\sqrt{(C_J+C_c)^2+(\omega{C}_cC_JZ_0/2)^2}},\\
&\Omega=\frac{C_c}{C_J+C_c-i\omega_dC_cC_JZ_0/2}\,
\frac{V_d}{\sqrt{2\hbar{Z}_J}},\\
&\tilde{\mathcal{T}}_{\mathbf{p}\mathbf{p}'}=
\sqrt{\frac{\hbar\omega_{p}}{8E_J}}\,
\mathcal{T}_{\mathbf{p}\mathbf{p}'}.
\end{align}\end{subequations}

The master equation is obtained assuming the following Ansatz for the density matrix of the full system (the TL, the JJ, and the quasiparticle) to hold at all times \citep{Breuer,Esposito2003,Lutchyn2006}:
\begin{equation}
\hat\rho^\mathrm{full}(t)=\sum_\mathbf{p}\hat\rho_\mathbf{p}(t)\otimes
\frac{|\mathrm{l},\mathbf{p}\rangle\langle\mathrm{l},\mathbf{p}|
+|\mathrm{u},\mathbf{p}\rangle\langle\mathrm{u},\mathbf{p}|}2
\otimes\hat\rho_\mathrm{tl}.
\end{equation}
Here $\hat\rho_\mathrm{tl}$ is the density matrix of the TL which is treated as an infinite bath, so its state cannot be changed by interaction with a finite number of degrees of freedom. We assume $\hat\rho_\mathrm{tl}$ to be that of the vacuum state (as discussed in Sec.~\ref{ssec:drive}, the effect of the incident wave is incorporated into the driving term in the Hamiltonian).
The quasiparticle is assumed to be located on any of the two islands with equal probability~\cite{Elastic}, thus the density matrix of the subsystem ``JJ + quasiparticle'' is proportional to the unit matrix in the island index $j=\mathrm{u,l}$.
We also assume the density matrix to be diagonal in the quasiparticle momentum, thereby neglecting any coherence between different quasiparticle states (this assumption is discussed in more detail in the end of this section). 
Thus, $\hat\rho_\mathbf{p}$ is the non-trivial part of the system density matrix which remains after having factored out the vacuum $\hat\rho_\mathrm{tl}$ and the unit matrix in the island index.

The subsequent steps are quite standard. Passing to the interaction representation,
\begin{align*}
&\hat\rho^\mathrm{full}(t)=e^{-i\hat{H}_0t/\hbar}\,
\hat{\tilde{\rho}}^\mathrm{full}(t)\,e^{i\hat{H}_0t/\hbar},\\
&\hat{\tilde{H}}_1(t)=
e^{-i\hat{H}_0t/\hbar}\,\hat{H}_1\,e^{i\hat{H}_0t/\hbar},
\end{align*}
and treating $\hat{H}_1$ as a perturbation, we obtain the
equation for $\hat{\tilde{\rho}}_\mathbf{p}(t)$ as
\begin{align}
&\frac{d\hat{\tilde{\rho}}_\mathbf{p}(t)}{dt} =
-\frac{1}{\hbar^2}\int\limits_{-\infty}^t{d}t'\times{}\nonumber\\
&\qquad{}\times
\sum_{j=\mathrm{u,l}}\left\langle{j},\mathbf{p}\left|\Tr_\mathrm{tl}\left\{
[\hat{\tilde{H}}_1(t),[\hat{\tilde{H}}_1(t'),
\hat{\tilde{\rho}}^\mathrm{full}(t')]]\right\}\right|{j},\mathbf{p}\right\rangle,
\label{eq:nonMarkovian}
\end{align}
where the trace is taken over the TL variables.
Using the Markovian approximation for the time integral,
neglecting fast oscillating terms, and going back to the original
Schr\"odinger representation, we arrive
at the following master equation for $\hat\rho_\mathbf{p}(t)$:
\begin{align}
&\frac{d\hat\rho_\mathbf{p}}{dt}=
-\frac{i}\hbar[\hat{H}_\mathrm{J},\hat\rho_\mathbf{p}]
+[\Omega{e}^{-i\omega_dt}\hat{a}^\dagger-\Omega^*e^{i\omega_dt}\hat{a}
,\hat\rho_\mathbf{p}]
+{}\nonumber\\ &\qquad\qquad{}+
\Gamma_\mathrm{tl}\,\hat{a}\hat\rho_\mathbf{p}\hat{a}^\dagger
-\frac{\Gamma_\mathrm{tl}}{2}\left\{\hat{a}^\dagger\hat{a},\hat\rho_\mathbf{p}\right\}
+{}\nonumber\\ &\qquad\qquad{}+
\frac{\omega_{p}\delta^2}{4\pi\Delta}\sum_\mathbf{p'}
\delta(\epsilon_\mathbf{p}-\hbar\omega_p-\epsilon_\mathbf{p'})\,
\hat{a}\hat\rho_\mathbf{p'}\hat{a}^\dagger
+{}\nonumber\\ &\qquad\qquad{}+
\frac{\omega_{p}\delta^2}{4\pi\Delta}\sum_\mathbf{p'}
\delta(\epsilon_\mathbf{p}+\hbar\omega_p-\epsilon_\mathbf{p'})\,
\hat{a}^\dagger\hat\rho_\mathbf{p'}\hat{a}-{}
\nonumber\\ &\qquad\qquad{}-
\frac{\omega_{p}\delta^2}{8\pi\Delta}\sum_\mathbf{p'}
\delta(\epsilon_\mathbf{p}+\hbar\omega_p-\epsilon_\mathbf{p'})\,
\left\{\hat{a}^\dagger\hat{a},\hat\rho_\mathbf{p}\right\}-{}
\nonumber\\ &\qquad\qquad{}-
\frac{\omega_{p}\delta^2}{8\pi\Delta}\sum_\mathbf{p'}
\delta(\epsilon_\mathbf{p}-\hbar\omega_p-\epsilon_\mathbf{p'})\,
\left\{\hat{a}\hat{a}^\dagger,\hat\rho_\mathbf{p}\right\},
\end{align}
where $\Gamma_\mathrm{tl}$ is the JJ excitation decay rate due to emission of TL photons in the weak-coupling limit:
\begin{equation}\label{Gammatl=}
\Gamma_\mathrm{tl}=\frac{\omega_{p}(Z_0/2Z_J)C_c^2}{(C_J+C_c)^2+(Z_0/2Z_J)^2C_c^2}
\approx\frac{C_c}{C_J}\,\omega_p^2\tau_c.
\end{equation}
Since $\hat\rho_\mathbf{p}$ can depend on $\mathbf{p}$ only via energy~$\epsilon_\mathbf{p}$, it is convenient to pass to
\begin{equation}
\hat\rho(\epsilon,t)=\frac{1}{\nu(\epsilon)}\sum_\mathbf{p}\hat\rho_\mathbf{p}(t)\,
\delta\!\left(\frac{\xi_\mathbf{p}^2}{2\Delta}-\epsilon\right),
\end{equation}
with the normalization $\int\Tr\hat\rho(\epsilon)\,\nu(\epsilon)\,d\epsilon=1$,
where $\nu(\epsilon)\approx(1/\delta)\sqrt{2\Delta/\epsilon}$ is the quasiparticle density of states, defined in Eq.~(\ref{nu=}) above.
Then the master equation takes the form
\begin{align}
\label{mastereq}
&\frac{\partial\hat\rho(\epsilon)}{\partial{t}}=
-\frac{i}\hbar[\hat{H}_\mathrm{J},\hat\rho(\epsilon)]
+[\Omega{e}^{-i\omega_dt}\hat{a}^\dagger-\Omega^*e^{i\omega_dt}\hat{a}
,\hat\rho(\epsilon)]
+{}\nonumber\\ &\qquad\qquad{}+
\Gamma_\mathrm{tl}\,\hat{a}\,\hat\rho(\epsilon)\,\hat{a}^\dagger
-\frac{\Gamma_\mathrm{tl}}{2}\left\{\hat{a}^\dagger\hat{a},\hat\rho(\epsilon)\right\}
+{}\nonumber\\ &\qquad\qquad{}
+\frac{\omega_{p}\delta}{4\pi\Delta}
\sqrt{\frac{2\Delta}{\epsilon-\hbar\omega_{p}}}\,
\hat{a}\,\hat\rho(\epsilon-\hbar\omega_{p})\,\hat{a}^\dagger
+{}\nonumber\\ &\qquad\qquad{}
+\frac{\omega_{p}\delta}{4\pi\Delta}
\sqrt{\frac{2\Delta}{\epsilon+\hbar\omega_{p}}}\,
\hat{a}^\dagger\hat\rho(\epsilon+\hbar\omega_{p})\,\hat{a}-{}
\nonumber\\ &\qquad\qquad{}
-\frac{\omega_{p}\delta}{8\pi\Delta}
\sqrt{\frac{2\Delta}{\epsilon+\hbar\omega_{p}}}
\left\{\hat{a}^\dagger\hat{a},\hat\rho(\epsilon)\right\}-{}
\nonumber\\ &\qquad\qquad{}
-\frac{\omega_{p}\delta}{8\pi\Delta}
\sqrt{\frac{2\Delta}{\epsilon-\hbar\omega_{p}}}
\left\{\hat{a}\hat{a}^\dagger,\hat\rho(\epsilon)\right\},
\end{align}
where the square roots should be set to zero if the argument is negative (which may occur for $\epsilon<\hbar\omega_{p}$). In the next section, we will us Eq.~(\ref{mastereq}) to study the JJ dynamics in the presence of the quasiparticle.

To conclude this section, let us discuss the assumptions made in the derivation of Eq.~(\ref{mastereq}). Using the Markovian approximation for the time integral in Eq.~(\ref{eq:nonMarkovian}) is equivalent to calculating the transition rates in Eq.~(\ref{mastereq}) from the Fermi Golden Rule. In both cases, it is important that the energy spectrum of the final states for the transition is continuous, or at least discrete but sufficiently dense, so that the level spacing is smaller than the obtained transition rate. For the photon emission into the TL this is perfectly valid because the TL photon spectrum is continuous. However, the quasiparticle levels are discrete, and the relevant energy level spacing is $\sim\delta\sqrt{\max\{\epsilon,\hbar\omega_p\}/\Delta}$, which should be compared to the typical rate, $\sim\delta(\omega_p/\Delta)\sqrt{\Delta/\max\{\epsilon,\hbar\omega_p\}}$, from Eq.~(\ref{mastereq}). It is easy to see that the rate is always smaller. Thus, for the above derivation to be valid, we need the TL-induced broadening, $\Gamma_\mathrm{tl}$, to be sufficiently strong compared to the level spacing and thus to the quasiparticle tunneling rate~\cite{Phonons,Elastic}.

The very same broadening mechanism that justifies the Markovian approximation, also enables us to neglect the quasiparticle coherence (the off-diagonal elements of the density matrix in the quasiparticle subspace). Indeed, when the quasiparticle performs a transition from a level with the energy $\epsilon$ into a bunch of levels with energies spread over an interval of width $\sim\hbar\Gamma_\mathrm{tl}$ around $\epsilon\pm\hbar\omega_p$, the off-diagonal terms beating at relative frequencies $\sim\Gamma_\mathrm{tl}$ have already dephased on the quasiparticle tunneling time scale.
Thus, the quasiparticle is treated as a ``mini-bath'', in the sense that the coherence is neglected, but change of the quasiparticle state by exciting or deexciting the JJ is accounted for~\cite{Esposito2003}.

So far, we did not assume the separability of the density matrix $\hat\rho(\epsilon)$ into a product of the JJ and quasiparticle matrices. However, the smallness of the quasiparticle tunneling rate with respect to the photon emission rate enables us to do so. Indeed, during the time the quasiparticle stays on one level, the JJ exchanges many photons with the TL and fully samples the allowed part of its Hilbert space. Thus, in the following we will use the separable form $\hat{\rho}(\epsilon)=\hat\rho_J\,f(\epsilon)$, where $\hat\rho_J$ is the JJ density matrix which does not depend on the quasiparticle energy, and $f(\epsilon)$ is the quasiparticle distribution function. Both are normalized:
\begin{equation}
\mathrm{Tr}\,\hat{\rho}_J=1,\quad
\int_0^\infty{f}(\epsilon)\,\nu(\epsilon)\,d\epsilon = 1.
\end{equation}

\section{Solution of the master equation}\label{sec:solution}

\subsection{The role of anharmonicity in the junction}

Let us expand the cosine term in Eq.~(\ref{Hr=}) to the fourth order:
\begin{equation}
\hat{H}_\mathrm{J}=\hbar\omega_p
\left(\hat{a}^\dagger\hat{a}+\frac{1}{2}\right)
-\frac{E_C}{12}\left(\hat{a}^\dagger+\hat{a}\right)^4.
\end{equation}
Since $E_C\equiv{e}^2/(2C_J)\ll\hbar\omega_p$, the last term produces an anharmonic correction to the JJ level energies,
$E_n=\hbar\omega_p(n+1/2)-(E_C/2)(n^2+n+1/2)$.
For not too large~$n$, the anharmonic correction to the transition energy $E_{n+1}-E_n$ is small compared to $\hbar\omega_p$. However, we are studying a resonantly driven junction, so we are interested in drive frequencies $\omega_d$ close to the transition frequency,
\begin{equation}\label{resonance=}
|E_{n+1}-E_n-\hbar\omega_\mathrm{d}|\sim\hbar\Gamma_\mathrm{tl}.
\end{equation}
Then, even though $E_C\ll\hbar\omega_p$, the difference in energies of the first two transitions, $(E_2-E_1)-(E_1-E_0)=-E_C$, can be large compared to $\Gamma_\mathrm{tl}$, if $E_C\gg\hbar\Gamma_\mathrm{tl}$. In this case the resonance condition~(\ref{resonance=}) can be satisfied only for one of the transitions, so the JJ effectively behaves as a two-level system, also known as the transmon qubit~\cite{Koch2007}. In the opposite limit, $\hbar\Gamma_\mathrm{tl}\gg{E}_C$, the JJ can be treated as a harmonic oscillator, provided that its degree of excitation is not too high. Below we will consider both these limits separately. The qubit limit will be treated by simply truncating the JJ Hilbert space to two levels and by replacing the creation and annihilation operators $\hat{a}^\dagger,\hat{a}$ in the master equation~(\ref{mastereq}) by the Pauli raising and lowering counterparts, $\sigma_+,\sigma_-$.

\subsection{Effective quasiparticle temperature}

The kinetic equation for the quasiparticle distribution function $f(\epsilon)$ is obtained by taking the trace over the JJ variables in Eq.~(\ref{mastereq}):
\begin{align}
\frac{\partial{f}(\epsilon)}{\partial{t}}={}&{}
\frac{\omega_{p}\delta^2}{4\pi\Delta}\nu(\epsilon-\hbar\omega_p)
\left[\bar{n}f(\epsilon-\hbar\omega_p)
-(1\mp\bar{n})f(\epsilon)\right]+{}\nonumber\\
{}&{}+\frac{\omega_{p}\delta^2}{4\pi\Delta}
\nu(\epsilon+\hbar\omega_p)
\left[(1\mp\bar{n}) f(\epsilon+\hbar\omega_p)
- \bar{n}f(\epsilon)\right],
\label{kinetic=}
\end{align}
where $\bar{n}\equiv\Tr\{\sigma_+\sigma_-\hat\rho_J\}$ or $\bar{n}\equiv\Tr\{\hat{a}^\dagger\hat{a}\hat\rho_J\}$ is the average number of excitations in the JJ in the qubit or harmonic limit, respectively, and the upper/lower sign corresponds to the qubit/harmonic limit.

We are interested in the stationary situation, so we assume
$\bar{n}$ to be constant.
Then, the stationary solution of the kinetic equation~(\ref{kinetic=}) is~\cite{Flatten}
\begin{equation}\label{fBoltzmann=}
f(\epsilon) = \frac\delta{\sqrt{2\pi{T}_J\Delta}}\,
e^{-\epsilon/T_J}\,\theta(\epsilon),
\end{equation}
where $\theta(\epsilon)$ is the step function, and we defined
\begin{equation}\label{Teff=}
T_J\equiv\frac{\hbar\omega_p}{\ln(1/\bar{n}\mp{1})},
\end{equation}
which has the meaning of the JJ effective temperature.
We emphasize that this is just a convenient notation;
the JJ is \emph{not} in a thermal state~\cite{Louisell1973}.

\subsection{The junction state}

To find the JJ state, we multiply the master equation~(\ref{mastereq}) by $\nu(\epsilon)$ and integrate with respect to~$\epsilon$, which gives the equation for the JJ density matrix~$\rho_J(t)$,
\begin{align}
\frac{\partial\hat\rho_J}{\partial{t}}={}&{}
-i\omega_{p}[\hat{a}^\dagger\hat{a},\hat\rho_J]
+[\Omega{e}^{-i\omega_dt}\hat{a}^\dagger-\Omega^*e^{i\omega_dt}\hat{a},\hat\rho_J]
+{}\nonumber\\ &{}+(\Gamma_\mathrm{tl}+\Gamma^-_\mathrm{qp})\,\hat{a}\hat\rho_J\hat{a}^\dagger
-\frac{\Gamma_\mathrm{tl}+\Gamma^-_\mathrm{qp}}{2}\left\{\hat{a}^\dagger\hat{a},\hat\rho_J\right\}
+{}\nonumber\\ &{}+\Gamma^+_\mathrm{qp}\,\hat{a}^\dagger\hat\rho_J\hat{a}
-\frac{\Gamma^+_\mathrm{qp}}{2}\left\{\hat{a}\hat{a}^\dagger,\hat\rho_J\right\}+{}\nonumber\\
& {}+\Gamma_\phi^*\,
\hat{a}^\dagger\hat{a}\hat\rho_J\hat{a}^\dagger\hat{a}
-\frac{\Gamma_\phi^*}{2}\left\{\hat{a}^\dagger\hat{a}
\hat{a}^\dagger\hat{a},\hat\rho_J\right\}.
\label{mastereqres}
\end{align}
written here for the harmonic limit; in the qubit limit one should just replace $\hat{a}^\dagger\to\sigma_+$, $\hat{a}\to\sigma_-$, $\hat{a}^\dagger\hat{a}\to(\sigma_z+1)/2$. 
The last term in Eq.~(\ref{mastereqres}) represents a pure dephasing contribution with the rate $\Gamma_\phi^*$ that we included phenomenologically. Other dissipation mechanisms in the artificial atom can be straightforwardly incorporated into the master equation~(\ref{mastereqres}); in the weak-coupling approximation the corresponding rates should be simply added to the absorption, emission and dephasing terms of the equation.
In both qubit and harmonic limits, the rates $\Gamma^\mp_\mathrm{qp}$ of the JJ excitation/deexcitation by the quasiparticle are given by \begin{equation}\label{Gammampdef}
\Gamma^-_\mathrm{qp}=\frac{\omega_p}{2\pi}\int\limits_0^\infty
\frac{f(\epsilon)\,d\epsilon}{\sqrt{\epsilon(\epsilon+\hbar\omega_p)}},\quad
\Gamma^+_\mathrm{qp}=\frac{\omega_p}{2\pi}\int\limits_0^\infty
\frac{f(\epsilon+\hbar\omega_p)\,d\epsilon}{\sqrt{\epsilon(\epsilon+\hbar\omega_p)}}.
\end{equation}
These expressions are smaller than those in Ref.~\cite{Catelani2014} by a factor of~4; this is because in that reference equal occupation was assumed for each spin and at both sides of the junction, while here we have a single quasiparticle.

Equation~(\ref{mastereqres}) has the form of the standard master equation for a driven harmonic oscillator (or a driven two-level system in the qubit limit) coupled to a Markovian bath~\cite{Louisell1973}. The difference from the standard case is that the rates $\Gamma^\mp_\mathrm{qp}$ depend on~$f(\epsilon)$, which, in turn, depends on $\hat\rho_J$ itself. For the stationary distribution function~(\ref{fBoltzmann=}), they are given by
\begin{equation}\label{Gamma=Bessel}
\Gamma^\mp_\mathrm{qp}=
\Gamma_0\,e^{\pm\chi}\sqrt{2\chi/\pi}\, K_0(\chi),
\end{equation}
where $K_0(\chi)$ is the modified Bessel function, and
\begin{equation}\label{Gamma0=}
\Gamma_0\equiv\frac{\delta}{2\pi\hbar}
\sqrt{\frac{\hbar\omega_p}{2\Delta}}, \quad
\chi\equiv\frac{\hbar\omega_p}{2T_J}.
\end{equation}
These expressions are different from those obtained in Ref.~\cite{Catelani2014} under the assumption of equilibrium-like form for the distribution function: the difference stems from the fact that here the quasiparticle number is fixed to one. Also, we remind that here we assumed $\hbar\omega_p\ll\Delta$; corrections can be calculated analogously to Ref.~\cite{Catelani2014}.
The following asymptotic expressions for low and high temperature illustrate the overall behavior of the rates~$\Gamma^\mp_\mathrm{qp}$ from Eq.~(\ref{Gamma=Bessel}):
\begin{subequations}\begin{align}
&\Gamma^-_\mathrm{qp}\approx\Gamma_0,\quad
\Gamma^+_\mathrm{qp}\approx\Gamma_0e^{-\hbar\omega_p/T_J},
\quad T_J\ll\hbar\omega_p,\\
&\Gamma^\mp_\mathrm{qp}\approx\Gamma_0
\sqrt{\frac{\hbar\omega_p}{\pi{T}_J}}
\left(1\pm\frac{\hbar\omega_p}{2T_J}\right)
\ln\frac{4T_J}{e^\gamma\hbar\omega_p},
\quad T_J\gg\hbar\omega_p.
\end{align}\end{subequations}
Here $\gamma=0.577\ldots$ is the Euler-Mascheroni constant.

The stationary solution of Eq.~(\ref{mastereqres}), found in the standard way (namely, by rewriting it as Bloch equations in the qubit limit or by acting on it by $\hat{a}$ and tracing in the harmonic limit), determines the stationary oscillating coherent polarization in the two limits:
\begin{subequations}\begin{align}
&\langle\sigma_-\rangle=\frac{\Gamma_n}{\Gamma_\mathrm{tot}}\,
\frac{[\Gamma_\phi/2-i(\omega_p-\omega_d)]\,
\Omega{e}^{-i\omega_dt}}%
{(\omega_p-\omega_d)^2+\Gamma_\phi^2/4+2|\Omega|^2\Gamma_\phi/\Gamma_\mathrm{tot}},
\label{avsigma=}\\
&\langle\hat{a}\rangle=\frac{\Omega{e}^{-i\omega_dt}}%
{i(\omega_p-\omega_d)+(\Gamma_n+\Gamma_\phi^*)/2},\label{ava=}\\
&\Gamma_n\equiv
\Gamma_\mathrm{tl}+\Gamma^-_\mathrm{qp}-\Gamma^+_\mathrm{qp},\nonumber\\
&\Gamma_\mathrm{tot}\equiv
\Gamma_\mathrm{tl}+\Gamma^-_\mathrm{qp}+\Gamma^+_\mathrm{qp},
\quad \Gamma_\phi\equiv\Gamma_\mathrm{tot}+\Gamma_\phi^*.
\nonumber
\end{align}\end{subequations}
Equation~(\ref{avsigma=}) coincides with the textbook solution of the Bloch equations~\cite{Abragam} for $\Gamma_\mathrm{qp}^+=0$.
Here $\Gamma_n$ has the meaning of relaxation rate for the excitation number, and $\Gamma_\phi/2$ is the total dephasing rate.
The average number of excitations $\bar{n}$ is found straightforwardly in the two limits. It is more convenient to give an expression for $e^{\hbar\omega_p/T_J}$, which has the same form in both cases,
\begin{equation}\label{expomegaTeff=}
e^{\hbar\omega_p/T_J}=
\frac{\Gamma|\Omega|^2+(\Gamma_\mathrm{tl}+\Gamma^-_\mathrm{qp})
[(\omega_d-\omega_p)^2+\Gamma^2/4]}%
{\Gamma|\Omega|^2+\Gamma^+_\mathrm{qp}
[(\omega_d-\omega_p)^2+\Gamma^2/4]},
\end{equation}
provided that one substitutes $\Gamma=\Gamma_\mathrm{tot}$ in the qubit limit and $\Gamma=\Gamma_n$ in the harmonic limit.
Since the rates $\Gamma^\mp_\mathrm{qp}$ depend on~$T_J$, Eq.~(\ref{expomegaTeff=}) is a self-consistency equation for~$T_J$, strictly speaking. However, as $\hbar\omega_p/T_J$ changes from zero to infinity, the right-hand side of Eq.~(\ref{expomegaTeff=}) varies in an interval between two finite values, and this interval is very narrow for $\Gamma_0\ll\Gamma_\mathrm{tl}$. In fact, to find $T_J$ as a function of the drive amplitude~$V_d$ or of the incident power  $P_\mathrm{in}=2|V_d|^2/Z_0=2\hbar\omega_p|\Omega|^2/\Gamma_\mathrm{tl}$, one can simply neglect $\Gamma_\mathrm{qp}^\pm$ in Eq.~(\ref{expomegaTeff=}), whose right-hand side then becomes $1+[(\omega_d-\omega_p)^2+\Gamma_\mathrm{tl}^2/4]/|\Omega|^2$. Then, the relation between the input power and the effective temperature is
\begin{equation}\label{expomegaTeffsimple=}
e^{\hbar\omega_p/T_J}=1+
\frac{(\omega_d-\omega_p)^2+\Gamma_\mathrm{tl}^2/4}{\Gamma_\mathrm{tl}/2}
\,\frac{\hbar\omega_p}{P_\mathrm{in}}
\equiv{1}+\frac{P_*}{P_\mathrm{in}}.
\end{equation}

\subsection{Transmission coefficient and quality factors}

Equations (\ref{avsigma=}) and (\ref{ava=}), together with Eq.~(\ref{quanti}), determine the average $\langle{Q}_J(t)\rangle$,
which, when substituted into Eq.~(\ref{S21}), gives the following expressions for the transmission coefficient in the qubit and the harmonic limits, respectively:
\begin{subequations}\begin{align}
S_{21}(\omega) = 1 - {}&{}
\frac{i\Gamma_\mathrm{tl}}2\frac{\Gamma_n}{\Gamma_\mathrm{tot}}
\frac{\omega-\omega_p-i\Gamma_\phi/2}%
{(\omega-\omega_p)^2+\Gamma_\phi^2/4
+2|\Omega|^2\Gamma_\phi/\Gamma_\mathrm{tot}},
\label{S21qubit=}\\
S_{21}(\omega) = 1 - {}&{}
\frac{i\Gamma_\mathrm{tl}/2}{\omega-\omega_p
+i(\Gamma_n+\Gamma_\phi^*)/2},
\label{S21harmonic=}
\end{align}\end{subequations}
where we used Eq.~(\ref{Gammatl=}), as well as the weak-coupling
assumptions $C_c\ll{C}_J$ and $\omega_pC_cZ_0\ll{1}$. We also consider driving not too far from resonance, $|\omega_d -\omega_p| \ll \omega_p$.
For $\Gamma_\mathrm{qp}^\pm=0$, Eq.~(\ref{S21qubit=}) coincides with Eq.~(55) of Ref.~\citep{Peropadre2013}.
Eq.~(\ref{S21harmonic=}) can be compared to the
phenomenological expression for the transmission coefficient near
a resonance \cite{Khalil2012,Megrant2012,Geerlings2012,Dumur2015}:
\begin{equation}\label{S21phenomenological=}
S_{21}(\omega)=S_{21}^\infty\,
\frac{\omega-\omega_0+i\omega_0/(2\mathcal{Q}_\mathrm{i})}%
{\omega-\omega_\infty+i\omega_0/(2\mathcal{Q}_\mathrm{i})
+i\omega_0/(2\mathcal{Q}_\mathrm{e})}.
\end{equation}
Here $S_{21}^\infty$ is the constant high-frequency asymptote,
$\omega_\infty$ and $\omega_0$ are the resonant frequencies
with and without coupling to the TL (we neglected
the frequency shifts in our weak-coupling limit, so both coincide
with the plasma frequency~$\omega_p$), and $\mathcal{Q}_\mathrm{e}$
and $\mathcal{Q}_\mathrm{i}$ are the external and internal quality
factors, respectively.
Thus, we adopt the following expressions for the external and internal quality factors in terms of the quantities, calculated above:
\begin{align}
&\mathcal{Q}_\mathrm{e} = \frac{\omega_p}{\Gamma_\mathrm{tl}}\approx
\frac{2Z_J}{Z_0}\,\frac{C_J^2}{C_c^2},\quad
\mathcal{Q}_\mathrm{i}=\frac{\omega_p}{\Gamma_\phi^*+\Gamma^-_\mathrm{qp}+\Gamma^+_\mathrm{qp}\pm 2\Gamma^+_\mathrm{qp}},\label{QeQi=}
\end{align}
where the upper/lower sign in the expression for $\mathcal{Q}_\mathrm{i}$ corresponds to the qubit/harmonic limit.

While Eq.~(\ref{QeQi=}) is straightforwardly obtained by comparing Eqs. (\ref{S21harmonic=}) and (\ref{S21phenomenological=}) in the harmonic limit, for the qubit limit Eq.~(\ref{S21qubit=}) can be cast into the form~(\ref{S21phenomenological=}) only when the power broadening term, $\propto|\Omega|^2$, in the denominator of Eq.~(\ref{S21qubit=}) is neglected.
The low-power condition reads
\begin{equation}
8 |\Omega|^2 \ll \Gamma_\mathrm{tot}\Gamma_\phi \sim\Gamma_\mathrm{tl}^2,
\end{equation}
which for near-resonant driving, $|\omega_d - \omega_p| \lesssim \Gamma_\mathrm{tl}$, is equivalent to $P_\mathrm{in} \ll P_*$. This condition then implies, via Eq.~(\ref{expomegaTeffsimple=}), low quasiparticle effective temperature, $T_J \ll \omega_p$.
Moreover, even in this regime, in order to attribute any power-dependent broadening to heating of quasiparticles, the more stringent condition,
\begin{equation}
4 |\Omega|^2 \ll \Gamma_\mathrm{tl} \left(\Gamma^-_\mathrm{qp}+\Gamma^+_\mathrm{qp}\right),
\end{equation}
should be met. Since already under the weaker condition we have $\Gamma^+_\mathrm{qp} \ll \Gamma^-_\mathrm{qp}$, and $\Gamma^-_\mathrm{qp} \approx \Gamma_0$ is independent of power, we conclude that no spectroscopic signature of quasiparticle heating can be detected in the qubit case.

In the above considerations, we have neglected the pure dephasing rate $\Gamma_\phi^*$. Experimentally, it can be made as small as $\Gamma_\phi^*\sim{10}$~kHz (see Ref.~\cite{Hutchings2017} and references therein for a recent discussion), a value comparable to $\Gamma_0$, see Table~\ref{tab:energiesJJ}. Theoretically, one can estimate the pure dephasing rate $\Gamma_\phi^*$ due to the quasiparticle as done in Ref.~\cite{Catelani2014}; it is shown there that such dephasing rate is of the order of the elastic tunneling rate $\Gamma_\mathrm{qp}^\mathrm{el}$~\cite{Elastic}. The latter, for realistic parameter values, is in principle power-dependent, but is at most comparable to $\Gamma_0$, see Tables~\ref{tab:paramsJJ} and~\ref{tab:energiesJJ}. Therefore, including $\Gamma_\phi^*$ does not change our conclusions for the qubit regime.

\begin{figure}
\includegraphics[width=8cm]{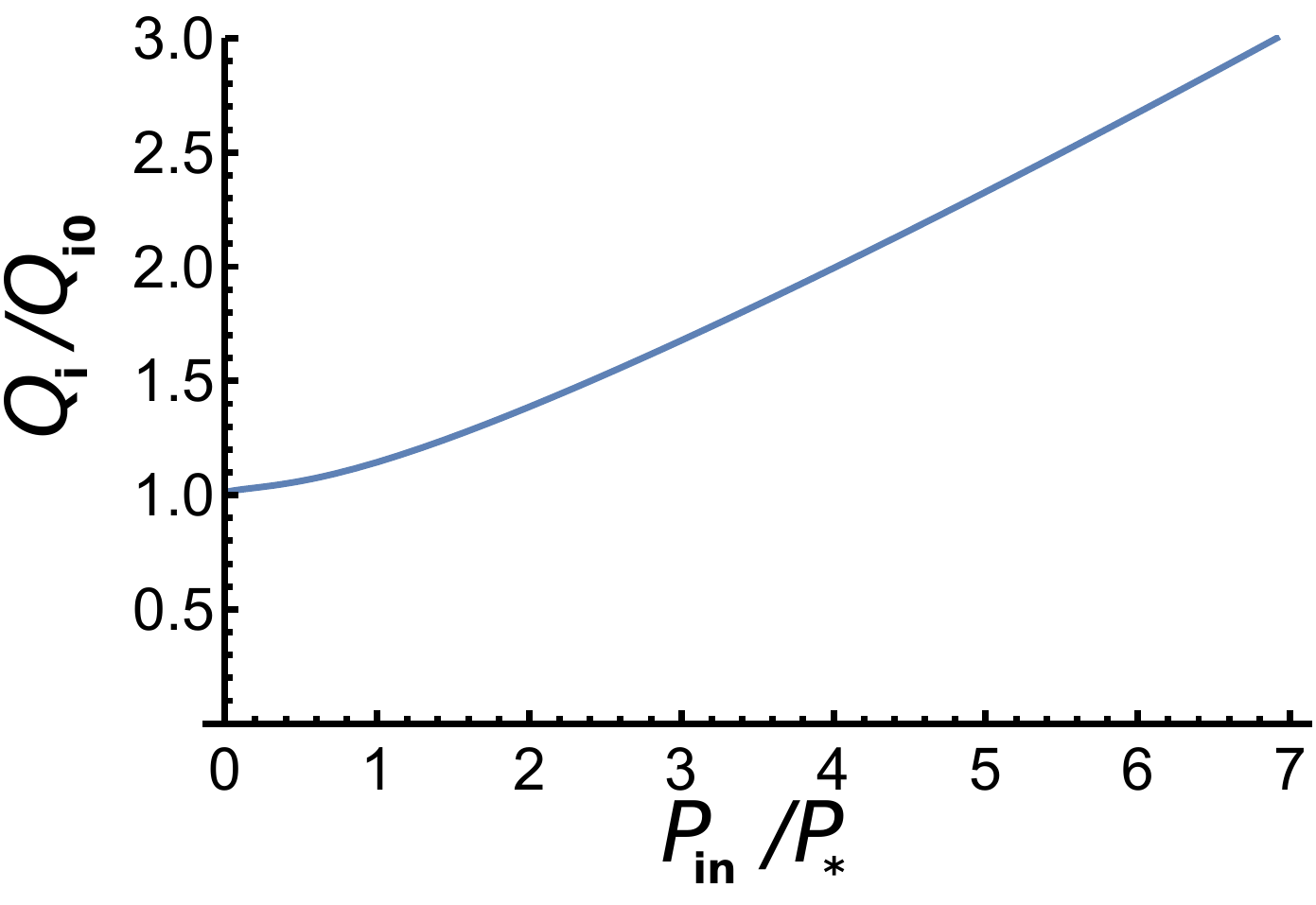}
\caption{\label{fig:Q_nophonons}
Plot of the internal quality factors $\mathcal{Q}_\mathrm{i}$, relative to the low-power value $\mathcal{Q}_\mathrm{i0}=\omega_p/\Gamma_0$, as a function of the dimensionless input power, $P_\mathrm{in}/P_*$, for the harmonic limit, neglecting the pure dephasing contribution.
}
\end{figure}

In Fig.~\ref{fig:Q_nophonons} we plot $\mathcal{Q}_\mathrm{i}$, relative to its low-power value,
$\mathcal{Q}_\mathrm{i0}\equiv\omega_p/\Gamma_0$,
as a function of the dimensionless input power, $P_\mathrm{in}/P_*$, as obtained from Eqs.~(\ref{Gamma=Bessel}) and (\ref{expomegaTeffsimple=}), for the harmonic limit and neglecting the pure dephasing $\Gamma_\phi^*$ for simplicity.
For a numerical estimate, we use typical structure parameters from Tables~\ref{tab:paramsJJ} and~\ref{tab:energiesJJ}.
However, the parameters in these tables show that the junction is in the qubit limit, $E_C\gg\hbar\Gamma_\mathrm{tl}$, so the quasiparticle heating effect is smeared by the power broadening. In the next section, we show that the harmonic limit is relevant for an artificial atom, represented by a chain of Josephson junctions.

\begin{table}
\begin{tabular}{|l|c|}
\hline TL impedance $Z_0$ & 50~$\Omega$\\
\hline JJ inductance $L_J$ & 1~nH \\
\hline JJ capacitance $C_J$ & 100~fF\\
\hline Coupling capacitance~$C_c$ & 10~fF\\
\hline Superconducting gap~$\Delta$ &
$200\:\mu\mbox{eV}=2\pi\hbar\times 48.4\:\mbox{GHz}$\\
\hline Density of states~$D_F$ &
$1.45\times 10^{47}\:\mbox{J}^{-1}\mbox{m}^{-3}$\\
\hline  Island volume & $0.1\:\mu\mbox{m}^3$\\
\hline  Electron-phonon coupling~$\Sigma$\; &
$2.0\times 10^8\:\mbox{W}\cdot\mbox{m}^{-3}\cdot\mbox{K}^{-5}$\\
\hline
\end{tabular}
\caption{\label{tab:paramsJJ}
Parameters for a Josephson junction coupled to a transmission line. Superconductor material parameters are taken for aluminum.}
\end{table}
\begin{table}
\begin{tabular}{|l|c|}
\hline Josephson energy~$E_J$ &
$680\:\mu\mbox{eV}=2\pi\hbar\times 160\:\mbox{GHz}$\\
\hline Charging energy $e^2/(2C_J)$ &
$0.80\:\mu\mbox{eV}=2\pi\hbar\times 194\:\mbox{MHz}$ \\
\hline Plasma frequency $\hbar\omega_p$ &
$66\:\mu\mbox{eV}=2\pi\hbar\times 16\:\mbox{GHz}$\\
\hline Mean level spacing $\delta$ &
$0.86\:\mbox{neV}=2\pi\hbar\times 0.21\:\mbox{MHz}$ \\
\hline Photon emission rate $\hbar\Gamma_\mathrm{tl}$ &
$136\:\mbox{neV}=2\pi\hbar\times 33\:\mbox{MHz}$\\
\hline Quasiparticle rate $\hbar\Gamma_0$
& $66\:\mbox{peV}=2\pi\hbar\times 13\:\mbox{kHz}$ \\
\hline Phonon rate $\hbar/\tau_\mathrm{ph}(\epsilon=\hbar\omega_p)$ &
$39\:\mbox{peV}=2\pi\hbar\times 9.5\:\mbox{kHz}$\\
\hline
\end{tabular}
\caption{\label{tab:energiesJJ} Energy scales for a Josephson junction coupled to a transmission line, derived from the parameters in Table~\ref{tab:paramsJJ}.}
\end{table}

\section{Few quasiparticles in a Josephson junction chain}
\label{sec:chain}

The theory developed in the previous sections can be quite straightforwardly extended to the case when the artificial atom is represented by a more complex system: instead of a single Josephson junction coupled to the transmission line, we consider a chain of $N-1$ junctions connecting $N$ islands. Each junction is characterized by the same parameters $E_J$ and $C_J$ as before, and, in addition each island is assumed to have a small capacitance $C_g\ll{C}_J$ to the ground. The first island is coupled to the TL via a capacitance~$C_c$. Such a chain has $N$ eigenmodes with frequencies~\cite{Pop2011,Masluk2012}
\begin{equation}
\omega_k=\omega_p\sqrt{\frac{\beta_kC_J}{C_g+\beta_kC_J}},\quad
\beta_k\equiv{2}-2\cos\frac{\pi{k}}N,
\end{equation}
where $k=0,\ldots,N-1$ and $\omega_p=1/\sqrt{L_JC_J}$ is the same plasma frequency as before. For sufficiently long chains, $N\gtrsim\ell_s\equiv\sqrt{C_J/C_g}$, the first few modes are well separated in frequency from each other and from the rest of the modes. Any of these first modes can be treated as a harmonic oscillator, and all the theory developed above for a single junction can be applied to this mode as well, with some modification of parameters. Namely, the rate of photon emission into the transmission line becomes
\begin{equation}
\Gamma_\mathrm{tl}=\frac{2}N\,\frac{\omega_k^2\tau_cC_c}{C_g+\beta_kC_J}.
\end{equation}
The anharmonic correction to the energy of mode~$k$ with $n_k$ photons can be written as $(\hbar{K}_{kk}/2)n_k^2$, with the self-Kerr coefficient ${K}_{kk}$ given by~\cite{Weissl2015}:
\begin{equation}\label{Kerr=}
\hbar{K}_{kk}=\frac{(\hbar\omega_k)^2}{2N^2E_J}\sum_{j=1}^{N-1}
\sin^4\frac{\pi{k}j}N.
\end{equation}
For $k=1$ and $N>2$ the sum is equal to $3N/8$. Numerical values of parameters for the lowest mode ($k=1$) of a chain of $N=201$ islands, given in Table~\ref{tab:JJchain}, show that the mode is in the harmonic limit, due to low frequency and large~$N$.

\begin{table}
\begin{tabular}{|l|c|}
\hline Island ground capacitance $C_g$ & 1~fF\\
\hline Chain length~$N$ & 201\\
\hline Lowest mode frequency $\hbar\omega_1$ &
$10.2\:\mu\mbox{eV}=2\pi\hbar\times 2.5\:\mbox{GHz}$\\
\hline Self-Kerr coefficient $\hbar{K}_{11}$ &
$0.14\:\mbox{neV}=2\pi\hbar\times 34\:\mbox{kHz}$\\
\hline Photon emission rate $\hbar\Gamma_\mathrm{tl}$ &
$3.8\:\mbox{neV}=2\pi\hbar\times 0.92\:\mbox{MHz}$\\
\hline Quasiparticle rate $\hbar\Gamma_0$
& $0.22\:\mbox{peV}=2\pi\hbar\times 53\:\mbox{Hz}$ \\
\hline Phonon rate $\hbar/\tau_\mathrm{ph}(\epsilon=\hbar\omega_1)$ &
$57\:\mbox{feV}=2\pi\hbar\times 14\:\mbox{Hz}$\\
\hline
\end{tabular}
\caption{\label{tab:JJchain}
Parameters and energy scales for the lowest mode of a 200-junction chain coupled to a transmission line.}
\end{table}

The quasiparticle state is now characterized by the island number $j=1,\ldots,N$, as well as momentum~$\mathbf{p}$. The quasiparticle can tunnel across any of the $N-1$ junctions, with the Hamiltonian
\begin{eqnarray}
&&\hat{H}_\mathrm{Jqp}=\sum_{j,\mathbf{p},\mathbf{p}'}
i\mathcal{T}_{\mathbf{p}\mathbf{p}'}^{(j)}
\left(\hat{a}_k+\hat{a}_k^\dagger\right)
\times{}\nonumber\\ && \qquad{}\times
\left(|j-1,\mathbf{p}\rangle\langle{j},\mathbf{p}'|
-|j,\mathbf{p}\rangle\langle{j-1},\mathbf{p}'|\right),\\
&&\mathcal{T}_{\mathbf{p}\mathbf{p}'}^{(j)}=
\sqrt{\frac{\hbar\omega_k}{4NE_J}}\,
\mathcal{T}_{\mathbf{p}\mathbf{p}'}\sin\frac{\pi{k}j}{N}.
\end{eqnarray}
We assume that the quasiparticle can reside on any island with equal probability, and can tunnel to any of the two neighboring islands.
Then the intrinsic quality factor is determined by
\begin{equation}
\Gamma_0=\frac\delta{N\pi\hbar}\,\sqrt{\frac{\hbar\omega_k}{2\Delta}},
\end{equation}
instead of Eq.~(\ref{Gamma0=}).
As seen from Table~\ref{tab:JJchain}, this rate is very low. However, a long chain should contain an extensive number $N_\mathrm{qp}$ of quasiparticles. As long as $N_\mathrm{qp}\ll{N}$, they can be treated independently, and their effect is additive, so $\Gamma_0$ should be multiplied by $N_\mathrm{qp}$.
Using the parameters from Table~\ref{tab:JJchain}, we obtain $\mathcal{Q}_\mathrm{i0}\approx{5}\times{10}^7/N_\mathrm{qp}$, which gives $\mathcal{Q}_\mathrm{i0}\approx{10}^6$ for $N_\mathrm{qp}=50$ (one quasiparticle per four junctions)
This value is rather high, so the quasiparticle-related dissipation will be important only if not masked by other mechanisms. Still, such high quality factors, varying from $10^6$ at low power to $10^7$ at high power, have been reported for superconducting resonators~\cite{Megrant2012}, and quality factors larger than $10^6$ have been obtained in superconducting qubits, both comprising one or two junctions (transmon~\cite{Wang2014,Hutchings2017}) or about 100 junctions (fluxonium ~\cite{Pop2014}).

\section{Quasiparticle relaxation by phonon emission}\label{sec:phonon}

In the above calculations we neglected the effect of the quasiparticle energy relaxation by phonon emission. To check the validity of this assumption, let us estimate the corresponding rate. For a quasiparticle with the energy much higher than the phonon temperature, the rate of acoustic phonon emission was calculated in Ref.~\cite{Kaplan1976}:
\begin{equation}\label{phononrelax}
\frac{1}{\tau_\mathrm{ph}(\epsilon)}=\frac{16}{315\,\zeta(5)}\,
\frac{\Sigma\epsilon^{7/2}}{D_F\sqrt{2\Delta}}.
\end{equation}
Here we introduced the effective coupling strength $\Sigma$, which controls energy exchange between electrons and phonons for the material in the normal state: the power per unit volume transferred from electrons and phonons which are kept at temperatures $T_\mathrm{e}$ and $T_\mathrm{ph}$, respectively, is given by $\Sigma(T_\mathrm{e}^5-T_\mathrm{ph}^5)$ \cite{Wellstood1994}. The coefficient $\Sigma$ can be represented in terms of the microscopic material parameters as
\begin{equation}
\Sigma=\frac{6\zeta(5)\,D_F\Xi^2}{\pi\hbar^4{v}_F\rho_0v_s^4},
\end{equation}
where $\Xi$ is the deformation potential, $v_F$ and $v_s$ are the Fermi velocity and the speed of sound, $\rho_0$~is the mass density of the material, $D_F$ is the density of states at the Fermi level for the material in the normal state, taken per unit volume and for both spin projections. The coefficient $\Sigma$ can also be measured experimentally (see Ref.~\cite{Giazotto2006} for a review). Using the parameters of aluminum, we estimate the phonon emission rate at energy $\epsilon=\hbar\omega_p$ for a single junction or $\epsilon=\hbar\omega_1$ for a Josephson junction chain (Tables~\ref{tab:energiesJJ} and~\ref{tab:JJchain}). The phonon emission rate is smaller than the quasiparticle tunneling rate, but the inequality is not very strong. Thus, a more detailed study of the competition between the two relaxation mechanisms is desirable. Such a study is beyond the scope of the present paper and will be the subject of a future publication~\cite{Catelani2018}.

\section{Conclusions}\label{sec:conclusions}

To conclude, we have studied intrinsic dissipation due to quasiparticle tunneling in a superconducting artificial atom, represented by a single Josephson junction or a Josephson junction chain. The artificial atom  is assumed to contain exactly one residual quasiparticle and is capacitively coupled to a coherently driven transmission line. In contrast to previous studies of quasiparticle-induced dissipation, we take into account heating of the quasiparticle by the drive. For simplicity, we assume that quasiparticle cooling by acoustic phonon emission is inefficient and can be neglected, so that the quasiparticle state is determined by the coupling to the superconducting degrees of freedom. We show that the corresponding intrinsic quality factor, as measured in a transmission experiment, increases with the drive power. This happens because the quasiparticle density of states decreases with the quasiparticle energy, so at stronger drive the quasiparticle tunneling is slower.

\section*{Acknowledgements}

We acknowledge support from the European Research Council (Grant No. 306731) and partial support by the EU under REA grant agreement CIG-618258.


\begin{thebibliography}{42}
\bibitem [{Pho()}]{Phonons}%
  \BibitemOpen
  \href@noop {} {}\bibinfo {note} {The quasiparticle levels are also broadened
  by the phonon emission. However, in the regime we are interested in, this
  broadening is not sufficient. Indeed, the large phonon emission rate
  $\Gamma_\mathrm{ph}\gg(\delta/\hbar)\sqrt{\max\{T_\mathrm{eff},\hbar\omega_p\}/\Delta}$
  would imply $\Gamma_\mathrm{ph}\gg\Gamma_\mathrm{qp}^\pm$. Then
  $T_\mathrm{eff}$ is determined by the phonon temperature and we are back to
  the regime of Refs.~\cite{Catelani2011a,Catelani2011b}. Here we are
  interested in $\Gamma_\mathrm{ph}\ll\Gamma_\mathrm{qp}^\pm$ or
  $\Gamma_\mathrm{ph}\sim\Gamma_\mathrm{qp}^\pm$, so the quasiparticle levels
  remain discrete.}\BibitemShut {Stop}%
\bibitem [{Ela()}]{Elastic}%
  \BibitemOpen
  \href@noop {} {}\bibinfo {note} {Besides tunneling accompanied by an AA
  transition, the quasiparticle can also tunnel elastically, without changing
  the AA state. The Golden Rule estimate for the rate of this tunneling is \[
  \Gamma_\mathrm{qp}^\mathrm{el}\sim\frac\delta\hbar\,\frac{E_J}\Delta\,
  \frac{T_\mathrm{eff}}\Delta\sqrt{\frac\Delta{T_\mathrm{eff}}}. \] When
  $E_J\gtrsim\Delta$ (good contact, which is quite a realistic situation), this
  rate does exceed the quasiparticle level spacing
  $\delta\sqrt{T_\mathrm{eff}/\Delta}$, so the quasiparticle does tunnel
  between the two islands. However, this elastic tunneling does not lead to
  level broadening. Indeed, the elastic coupling results in strong
  hybridization between the quasiparticle states on the two islands, but the
  exact eigenstates, although delocalized over the two islands, are still
  discrete, with the level spacing just twice smaller than on a single
  island.}\BibitemShut {Stop}%
\bibitem [{Fla()}]{Flatten}%
  \BibitemOpen
  \href@noop {} {}\bibinfo {note} {Strictly speaking, any distribution of the
  form $f(\epsilon)=u(\epsilon)\,e^{-\epsilon/T_J}\,\theta(\epsilon)$ with an
  arbitrary periodic function $u(\epsilon)=u(\epsilon+\hbar\omega_p)$, is a
  stationary solution of Eq.~(\ref{kinetic=}); however, since the energy
  conservation in absorption/emission of energy quantum $\hbar\omega_p$ is not
  exact, but only up to the level broadening, the periodic part $u(\epsilon)$
  will eventually flatten.}\BibitemShut {Stop}%
\end{thebibliography}
%

\end{document}